\documentclass[12pt]{article}
\def\la{\langle}
\def\ra{\rangle}

\def\k2av{\la k_T^2\ra}
\def\beq{\begin{equation}}
\def\eeq{\end{equation}}
\def\be{\begin{eqnarray}}
\def\ee{\end{eqnarray}}
\def\hs{\hat{s}}
\def\htm{\hat{t}}
\def\hu{\hat{u}}
\newcommand{\f}[2]{\frac{#1}{#2}}
\newcommand{\dd}{ {\textrm d}}

\begin{document} 

\title{Jet quenching as a probe of gluon plasma formation}

\author{
{G. Fai}$^1$\footnote{Presented 
at the International Europhysics Conference on High Energy Physics
(HEP2001), Budapest, Hungary, 12-18 July 2001. 
JHEP, PRHEP-hep2001/242.}, 
G.G. Barnaf\"oldi$^2$, M. Gyulassy$^3$, 
P. L\'evai$^2$, \\
G. Papp$^4$, I. Vitev$^3$, and Y. Zhang$^1$ \\[1ex]
$^1$ CNR, Dept. of Physics, Kent State Univ., Kent, OH 44242, USA \\
$^2$ KFKI RMKI, P.O. Box 49, Budapest 1525, Hungary \\
$^3$ Physics Department, Columbia Univ., \\
538 W. 120-th Street, New York, NY 10027, USA \\
$^4$ HAS Research Group for Theoretical Physics, \\
P.O. Box 32, Budapest 1518, Hungary  \\
E-mail: {fai@cnrred.kent.edu} 
}

\date{November 16, 2001}

\maketitle

\begin{abstract}
We study, in a pQCD calculation augmented by nuclear effects,
the jet energy loss needed to
reproduce the $\pi^0$ spectra in Au+Au collisions at large $p_T$,
measured by PHENIX at RHIC. The transverse width of the parton momentum 
distributions ({\it intrinsic $k_T$})
is used phenomenologically to obtain a reliable baseline $pp$ result.
Jet quenching is applied to the nuclear spectra (including shadowing 
and multiscattering) to fit the data. 
\end{abstract}

\vspace*{-0.4 cm}
\section{Introduction}

The average energy loss of jets in central $Au+Au$ collisions at RHIC
can be used to estimate the density of gluons in the medium. To determine 
the effect of jet energy loss from the data,
one needs a reliable background calculation of pion and kaon 
spectra subject to jet quenching. Here we provide
the required  baseline pQCD calculation augmented 
by the nuclear effects of multiscattering and shadowing. 

\section{Parton model and pQCD with intrinsic transverse momentum}

The invariant cross section for the production of hadron $h$ in a $pp$ collision 
is described in the pQCD-improved parton model on the basis of the factorization 
theorem as a convolution\cite{FF95}: 
\beq
\label{hadX}
  E_{h}\f{\dd \sigma_h^{pp}}{\dd ^3p} =
        \sum_{abcd} \int\! \dd x_{a,b} \ \dd z_c \ f_{a/p}(x_a,k_{Ta},Q^2)\
        f_{b/p}(x_b,k_{Tb},Q^2)\ 
\f{\dd \sigma}{\dd \htm}\,
   \frac{D_{h/c}(z_c, Q'^2)}{\pi z_c^2} \, \hs \, \delta(\hs+\htm+\hu)\ , 
\eeq
where  $f_{a,b/p}(x,k_{T},Q^2)$
are the parton distribution functions (PDFs) for the
colliding partons $a$ and $b$ in the interacting protons 
as functions of momentum fraction $x$, at scale $Q$, 
incorporating a Gaussian transverse momentum distribution. Having convinced 
ourselves that it makes sense to use a $p_T$-independent width
at a given energy\cite{zfpbl01}, we treat the width of the
transverse momentum distribution, 
$\k2av_{pp}$, as an energy-dependent parameter (also see e.g. \cite{E609}).

In eq. (\ref{hadX}), $\dd \sigma/ \dd\htm$ is the hard scattering cross section of the
partonic subprocess $ab \to cd$, and $D_{h/c}(z_c, Q'^2)$, the fragmentation function (FF),
gives the probability for parton $c$ to fragment into hadron $h$
with momentum fraction $z_c$ at scale $Q'$.
The scales are fixed in the present work as $Q = p_T/2$
and $Q' = p_T/(2z_c)$. We use leading order (LO) partonic cross sections,
together with LO PDFs (GRV)\cite{GRV92} and FFs (KKP)\cite{KKP}. This ensures 
the consistency of the calculation. Approximate NLO treatments via a 
$p_T$ and energy independent ``K factor'' are also 
possible\cite{wang01b,PLF00}. However, the dependence of the 
K factor on $p_T$ and $\sqrt{s}$ can not be neglected if a wide range 
of these variables is to be addressed\cite{bflpz01}. In addition, we prefer to 
keep the effects of the K factor separate from those 
of the intrinsic transverse momentum.

Fig. 1 displays the extracted values of the width $\k2av_{pp}$
from hard pion production data at energies 
below $\sqrt{s} \leq$ 60 GeV and from
$h^\pm = (h^+ + h^-)/2$ data at higher energies. 
The solid curve is drawn to 
guide the eye, while the dashed lines delineate
an estimated band of increasing uncertainty with 
increasing $\sqrt{s}$. We use this band to 
read off the $\k2av_{pp}$ values to be applied at RHIC energies. 
The solid bars represent estimates at  $\sqrt{s}=$ 130 and 200 
GeV, respectively. We  use $\k2av_{pp}$= 2 GeV$^2$ in our pQCD calculations
analyzing recent RHIC results at $\sqrt{s}=$ 130 AGeV.

\section{Nuclear \ shadowing \ and \ multiscattering}

We approximately include the modification of the PDFs inside nuclei 
\mbox{(``shadowing'')} 
and the iso\-spin asymmetry of heavy nuclei into the nuclear 
PDFs via the average nuclear dependence and a 
scale independent shadowing function $S_{a/A}(x)$ 
adopted from 
\cite{wang91}:
\beq
\label{shadow}
f_{a/A}(x,Q^2) = S_{a/A}(x) \left[\frac{Z}{A} f_{a/p}(x,Q^2) + 
  \left(1-\frac{Z}{A}\right)
  f_{a/n}(x,Q^2) \right]   \,\,\,\,  ,
\eeq
where $f_{a/n}(x,Q^2)$ is the PDF for the neutron.

The enhancement of the width of the transverse momentum 
distribution of partons due to interactions in the medium 
(``multiscattering'') is treated in the Glauber framework\cite{glauber59}.
We hold multiscattering
accountable for the Cronin effect in $pA$ collisions, and use $pA$ data
to parameterize the width enhancement. The enhanced width is written as  
\beq
\label{ktbroadpA}
\k2av_{pA} = \k2av_{pp} + C \cdot (\nu_{A}(b)-1) \ ,
\eeq
where $\k2av_{pp}$ is the width of the parton transverse momentum distribution 
in $pp$ collisions, $\nu_{A}(b)-1$ is the number of {\it effective} 
nucleon-nucleon (NN) collisions at impact parameter~$b$,
which impart an average transverse momentum squared $C$. 
In $pA$ reactions, where one of the hard-colliding partons
originates in a nucleon with additional NN collisions, we use the
$pp$ width from Fig. 1 for one of the colliding partons and the 
width (\ref{ktbroadpA}) for the other:
\beq
\label{pAX}
  E_{h}\f{\dd \sigma_{h}^{pA}}{ \dd ^3p} =
       \int \dd ^2b \,\, t_A(b)\,\, E_{h} \,
       \f{\dd \sigma_{h}^{pp}(\k2av_{pA},\k2av_{pp})}
{\dd ^3p}  
\,\,\, ,
\eeq
where $t_A(b) = \int \dd z \, \rho(b,z)$ 
is the nuclear thickness function
normalized as $\int \dd ^2b \, t_A(b) = A$. 
The thickness function determines the collision number $\nu_A(b)$ as
$\nu_A (b) = \sigma_{NN} t_A(b)$, where $\sigma_{NN}$ is the inelastic
nucleon-nucleon cross section.
We obtain the best fit to the Cronin 
data if we maximize the number of effective NN collisions at  a given
impact parameter at $(\nu_{A}(b)-1)_{max}=3$, and take the value of the 
parameter $C=0.4$ GeV$^2$.

\section{Jet quenching}

In $AA$ reactions, where both hard-colliding partons 
originate in nucleons with additional semi-hard collisions, we use the
width (\ref{ktbroadpA}) for both initial partons. Thus,
\beq
\label{ABX}
  E_{h}\f{\dd \sigma_{h}^{AB}}{\dd ^3p} =
       \int \dd ^2b \, \dd ^2r \,\, t_A(r) \,\, t_B(|\vec b - \vec r|) \,
E_{h} \, \f{\dd \sigma_{h}^{pp}(\k2av_{pA},\k2av_{pB})}{\dd ^3p}  
\,\,\, ,
\eeq
with all parameters as fixed earlier. Applying this model to WA80\cite{WA80}
and WA98\cite{WA98} central $\pi^0$ production data at SPS energies, we find a 
reasonable agreement beyond a transverse momentum of $p_T \geq $ 2 -- 2.5 GeV,
with possibly an overestimation of the data by the model
in the heaviest system ($Pb+Pb$) by upto 40\%\cite{zfpbl01}. 
This may be taken as a hint that  
an additional mechanism is at work in the nuclear medium,
which acts to reduce the calculated cross sections. As we will see next, 
the discrepancy becomes more serious at RHIC energies. Since this effect 
can also be looked upon as a 
shift of the spectra to lower $p_T$, i.e. jet energy loss,
we study jet quenching\cite{plumer,baier98,gyulassy00}
as a potential explanation. 

Assuming a static gluon plasma, we use here the 
{\em average} energy loss of gluon jets to estimate the effect on the
cross sections. The fragmentation part of (\ref{hadX}) is modified as
\beq
\label{jetqX}
\f{D_{h/c}(z_c, Q'^2)}{\pi z_c^2}  \longrightarrow 
\f{z_c^*}{z_c} \f{D_{h/c}(z_c^*, Q'^2)}{\pi z_c^2} \ , 
\eeq
where the down-shifted (quenched) momentum $p_c^* = p_c - \Delta E$ results in a 
momentum fraction $z_c^* = z_c/\left( 1-\Delta E/p_c \right)$. 
For the energy shift $\Delta E$ we use the ``thin plasma'' (GLV)
approximation\cite{gyulassy00},
appropriate for small values of the opacity $\bar{n}=L/\lambda$, where $L$ is 
the average thickness of the static medium and $\lambda$ denotes the gluon 
mean free path. One important feature
distinguishing the GLV energy loss from the ``thick plasma'' 
(BDMS) limit\cite{baier98} is the approximately constant 
fractional energy loss ($\Delta E/E$) of gluons as a 
function of jet energy $E$ in the range 
$3 \leq E \leq 10$ GeV\cite{gyulassy00}.

In Fig. 2. the results of this calculation are compared 
to recent $\pi^0$ spectra measured by PHENIX \cite{david01,phenix01} 
for values of the opacity $0 \leq \bar{n} \leq 4$. 
It can be seen that, while the semi-peripheral spectrum displays only a
weak (but non-vanishing) influence of jet-quenching characterized by
$\bar{n} \approx$ 1 -- 1.5, in the central case
a much larger effective opacity of $\bar{n} \approx$ 3 -- 4 
is required for agreement between the data and pQCD calculations \cite{lpqm01}. 
It should be noted that fluctuations in jet energy loss reduce 
the suppression, but their net effect can be absorbed in a
rescaling of the opacity $\bar{n}$. First results on an approximate treatment
of fluctuations and on a dynamical calcula\-tion are reported in the  next   
talk\cite{vitev01}.

The average opacity $\bar{n}$ can be related to the density of the plasma 
using the expression of the transport coefficient in terms of 
$\dd N_g/\dd y$ \cite{gyulassy00}. Using this correspondence implies
a very high initial gluon density for $\bar{n} \approx$ 3 in a static plasma. 
Taking into account the rapid transverse collective expansion of the plasma 
leads to somewhat reduced gluon densities, $\dd N_g/\dd y \approx 500$,
and to results on the collective flow consistent with the 
preliminary data\cite{gyulassy01}. By any reasonable estimate these
gluon densities indicate that strongly-interacting matter has reached the 
deconfined region in the analyzed nuclear collisions.

\section{Summary and conclusions}

Jet quenching is needed on the background of a pQCD calculation augmented
by the nuclear effects of shadowing and multiscattering to bring the 
calculated results in agreement with the PHENIX data 
on $\pi^0$ spectra from $Au+Au$ at $\sqrt{s}=$ 130 GeV. The amount of required 
jet energy loss indicates a dense gluon plasma, well in the deconfined region.
Jet quenching is a powerful diagnostic tool to study strongly-interacting 
matter in nuclear collisions, which will play an important role in 
analyzing the 2001 RHIC data at $\sqrt{s}=$ 200 GeV, and beyond.

\bigskip

\section*{Acknowledgment:}

We thank G. David for discussions.
This work was supported in part by  U.S. NSF grant INT-0000211 (MTA-OTKA-NSF),
DOE grant DE-FG02-86ER40251,  and Hungarian grants FKFP220/2000, OTKA-T032796,
OTKA-T034269 and OTKA-T034842.

\newpage

\begin{center}
\vspace*{18.0cm}
\includegraphics{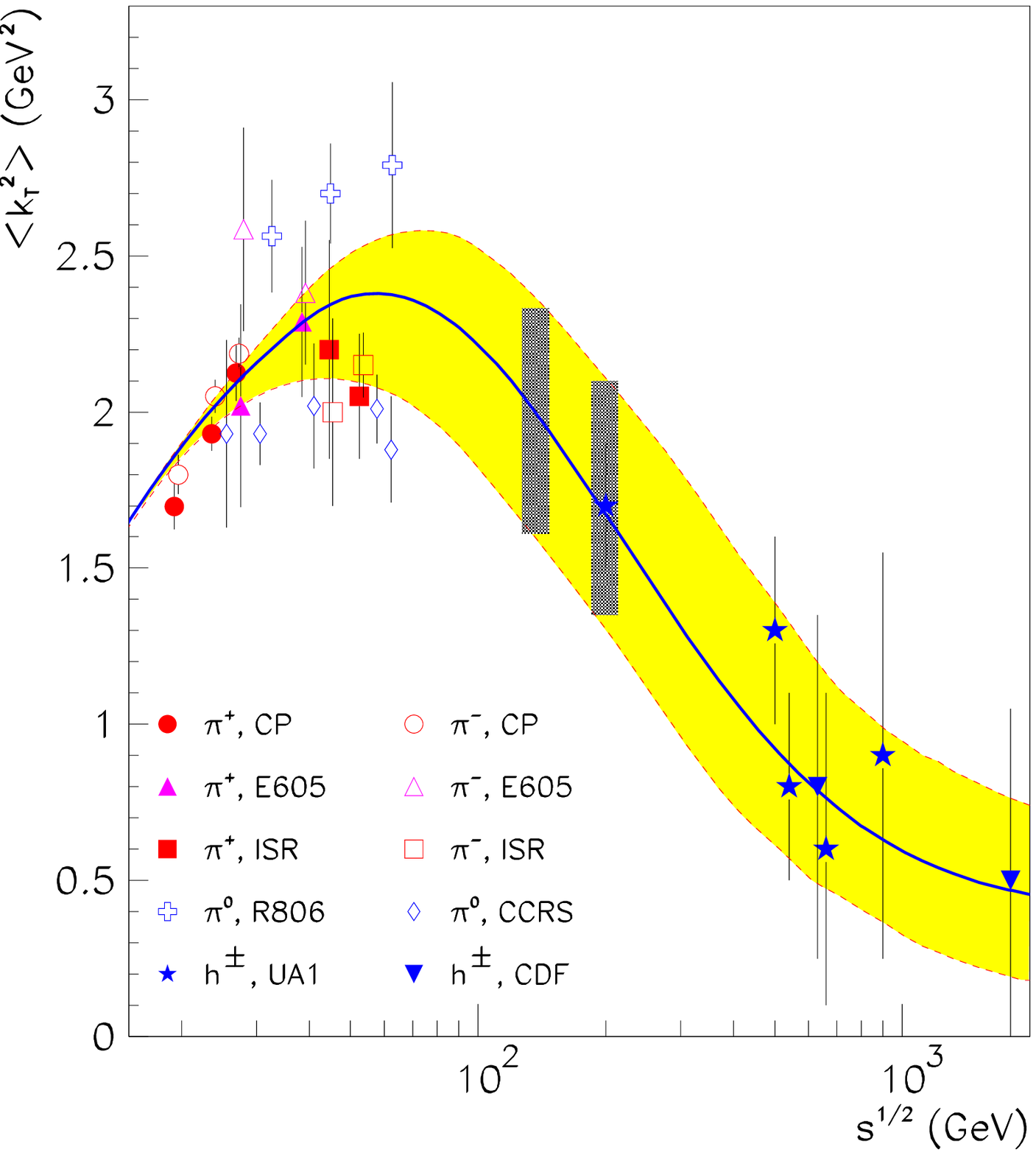}
\begin{minipage}[t]{15.cm}
{ {\bf Fig.~1.}
The best fit values of $\k2av_{pp}$ in $p p \rightarrow \pi X$ 
and $p \bar{p} \rightarrow h^\pm X$ reactions.
Where large error bars would overlap at the same energy,
one of the points has been shifted slightly for better 
visibility. The band is drawn to guide the eye.
}
\end{minipage}
\end{center}

\newpage
\begin{center}
\vspace*{18.0cm}
\includegraphics{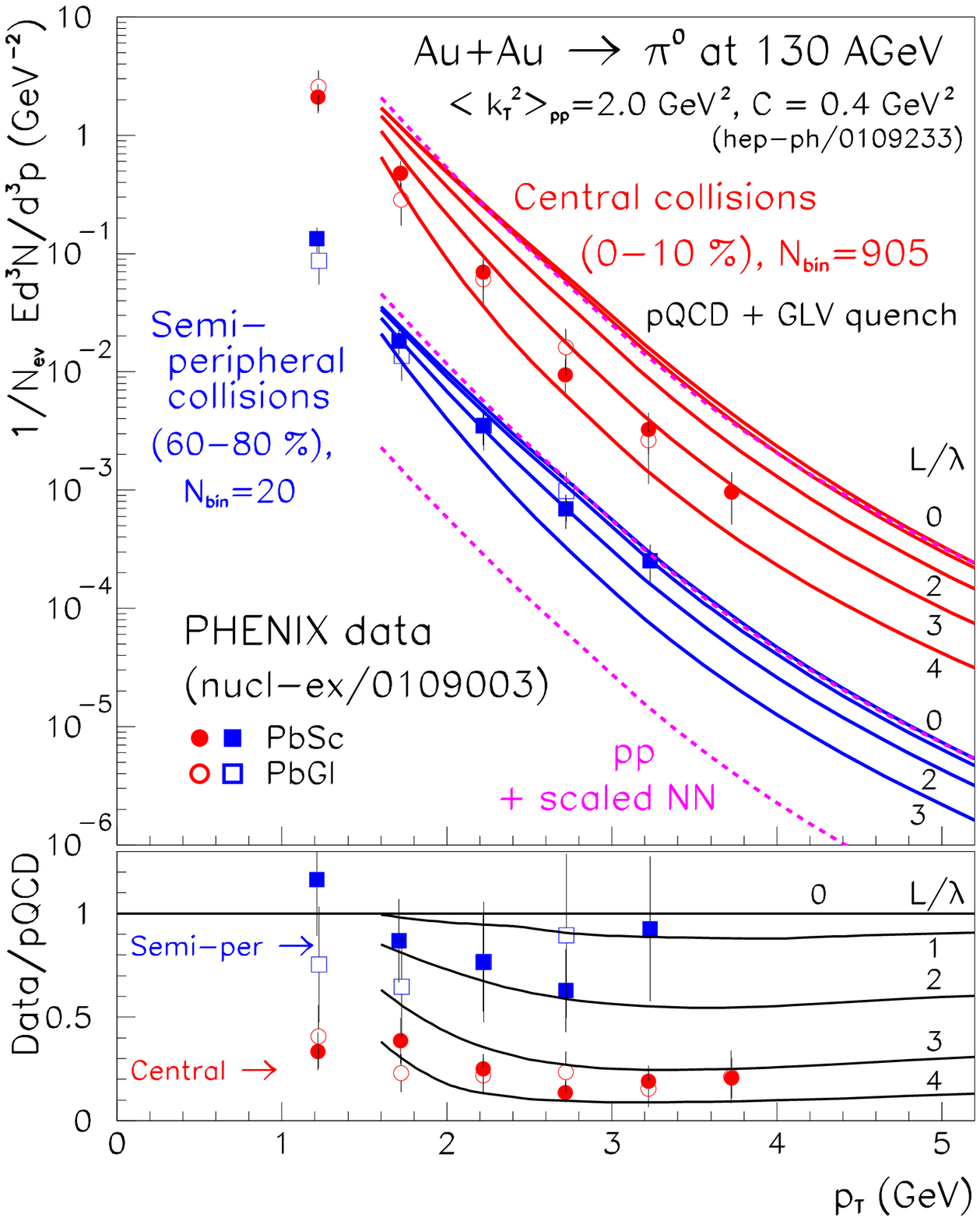}
\begin{minipage}[t]{15.cm}
{ {\bf Fig.~2.}
Semi-peripheral and central $\pi^0$ production in $Au+Au$ collisions at 
$\sqrt{s}=$ 130 GeV. Data are from Ref.~\cite{phenix01}. Calculation with 
jet quenching at opacities $L/\lambda =0,1,2,3,4$. Dashed lines are 
$pp\rightarrow \pi^0$ and Glauber-scaled estimates 
for Au+Au.
}
\end{minipage}
\end{center}

\end{document}